\documentclass[aps,prb,superscriptaddress,twocolumn]{revtex4-2}
\usepackage{amsmath,bm}
\usepackage{graphicx,epsfig,braket,amssymb,amsfonts,eufrak,tabularx,multirow,amsmath,graphicx,color}
\usepackage{newtxtext}
\usepackage[colorlinks=true]{hyperref}

\newcommand{\bd}{\begin{displaymath}}
\newcommand{\ed}{\end{displaymath}}
\renewcommand{\v}[1]{{\bf #1}}
\newcommand{\bpm}{\begin{pmatrix}}
\newcommand{\epm}{\end{pmatrix}}
\newcommand{\nn}{\nonumber \\}

\begin{document}

\title{Topological magnon-polarons in honeycomb antiferromagnets with spin-flop transition}

\author{Gyungchoon Go}
\affiliation{Department of Physics, Korea Advanced Institute of Science and Technology, Daejeon 34141, Korea}

\author{Heejun Yang}
\affiliation{Department of Physics and Astronomy, Seoul National University, Seoul 08826, South Korea}

\author{Je-Geun Park}
\affiliation{Department of Physics and Astronomy, Seoul National University, Seoul 08826, South Korea}

\author{Se Kwon Kim}
\affiliation{Department of Physics, Korea Advanced Institute of Science and Technology, Daejeon 34141, Korea}

\begin{abstract}
We theoretically investigate the thermal Hall transport of magnon-polarons in a two-dimensional honeycomb antiferromagnetic insulator under the influence of a perpendicular magnetic field, varying in strength. The application of a perpendicular magnetic field induces a magnetic phase transition from the collinear antiferromagnetic phase to the spin-flop phase, leading to a significant alteration in Hall transport across the transition point. In this paper, our focus is on the intrinsic contribution to thermal Hall transport arising from the magnetoelastic interaction. To facilitate experimental verification of our theoretical results, we present the dependence of thermal Hall conductivity on magnetic field strength and temperature.
\end{abstract}

\maketitle

\section{Introduction}

In recent years, intrinsic magnetism in two-dimensional (2D) insulators has been discorvered,
attracting growing attention, driven by its fundamental interest and technological applications in reduced dimensions~\cite{McGuire2015,Zhang2015,Park2016,Lee2016,Gong2017,Huang2017,Burch2018,Deng2018,Fei2018,OHara2018,Gibertini2019,Kim2019}.
Within magnetic magnetic insulators, magnetic excitations (magnons) and lattice vibrations (phonons) serve as carriers of energy and information.
The exploration of these collective, charge-neutral, low-energy excitations has attracted considerable interest due to their potential for innovative approaches to manipulate and control thermal energy and information~\cite{Chumak2015,Maldovan2013}.
A particular focus lies on the topological Hall transport of collective excitations resulting from Berry curvature.
Previous research has demonstrated the magnon Hall effect in chiral magnetic systems with chiral spin texture~\cite{Katsura2010,Han2017,Gobel2017,Diaz2019,Kim2019a}, or Dzyaloshiskii-Moriya interaction~\cite{Onose2010,Matsumoto2011,Zhang2013,Mook2014,Kim2016a,Cheng2016,Owerre2016,Zyuzin2016,Li2020}.
Additionally, studies have explored the phonon Hall effect arising from the interaction between phonons and static magnetization~\cite{Sheng2006,Kagan2008,Zhang2010},
or scattering from the impurities~\cite{Flebus2022,Sun2022,Guo2022}.
Beyond individual magnons and phonons, their hybrid excitations known as magnon-polarons potentially exhibit topological properties
through the long-range dipolar interaction~\cite{Takahashi2016}, the Dzyaloshiskii-Moriya interaction~\cite{Zhang2019,Park2020,Zhang2021,Bao2023},
and strain-dependent magnetic anisotropy~\cite{Go2019,Zhang2020,Shen2020a,Bazazzadeh2021,Huang2021,Go2022}.

The strain-dependent magnetic anisotropy is a well-known mechanism of magnetoelastic interaction and magnetostriction~\cite{Kittel1949,Kittel1958,Callen1965}, ubiquitous in magnetic materials.
Recent experiments have revealed its role in opening band gaps  between magnon and phonon modes~\cite{Liu2021,Luo2023} and inducing the topological transport of the quasiparticles~\cite{Zhang2021, Xu2023, Li2023}.
Most theoretical investigations into the magnetoelastic interaction-induced topological magnon-polaron have focused on collinear magnetic systems, where the magnetic order aligns in the uniform direction.
Consequently, they are not directly applicable to certain antiferromagnetic (AFM) systems, including the spin-flop ground state.
In the spin-flop phase of bipartite antiferromagnets, two spins on the two sublattices tilt at an angle $\theta_A = \theta_B$ (Fig.~\ref{fig:1}).
In the tilted ground state, magnons couple not only with out-of-plane phonons but also with in-plane phonons~\cite{Shen2020a}.
However, a theoretical investigation of topological magnon-polaron bands, including both in-plane and out-of-plane phonon modes in the AFM spin-flop state, is currently lacking.

In this paper, we investigate the topological properties of the magnon-polarons in a two-dimensional honeycomb antiferromagnet subjected to a variable magnetic field induced by the magnetoelastic interaction.
More specifically, we compute the band structures and topological properties of the magnon-polarons in both collinear AFM and spin-flop states.
To facilitate further comparison with experimental works, we calculate the magnetic field dependence of the thermal Hall conductivity at different temperatures.

The remaining sections of this paper are organized as follows.
In Sec.~II, we present model Hamiltonians for magnons, phonons, and magnetoelastic interaction in both collinear AFM and spin-flop states.
In Sec.~III, we calculate the band structures and Berry curvatures of the magnon-polaron bands.
In Sec.~IV, we compute the the thermal Hall conductivity of the magnon-polarons.
In Sec.~V, we conclude by providing a brief summary and discussion.

\section{Model Hamiltonian}

\subsection{Magnon part}

\begin{figure}[t!]
\includegraphics[width=0.9\columnwidth]{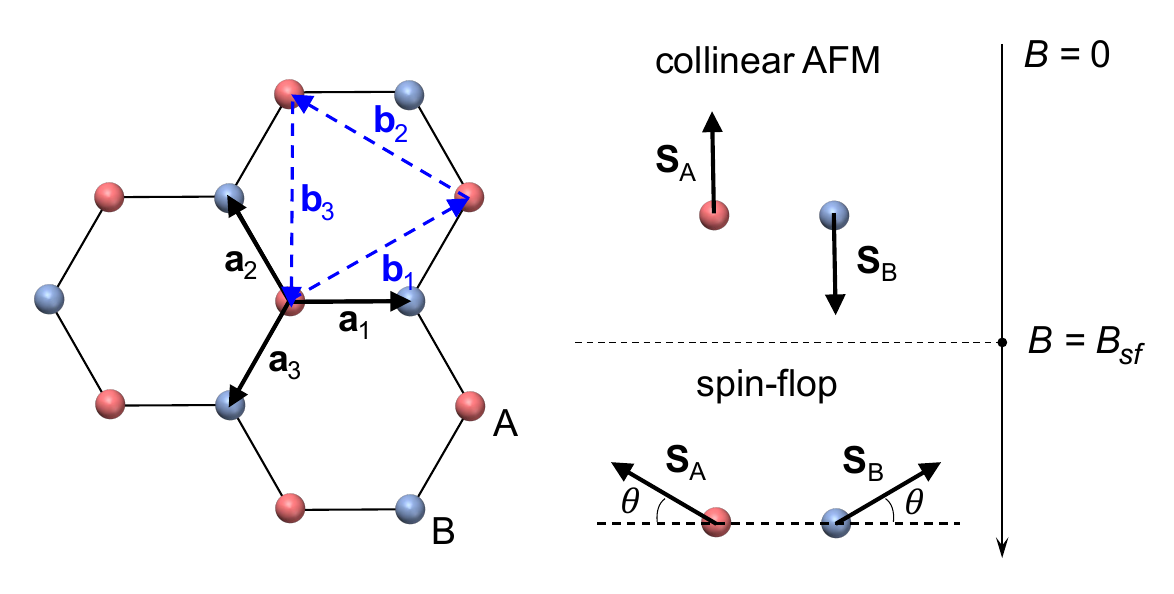}
\caption{Schematic illustration of the honeycomb lattice (left). Equilibrium spin configurations depending on external magnetic field (right).}\label{fig:1}
\end{figure}

Here, we consider the spin Hamiltonian for a 2D honeycomb AFM system, given by
\begin{align}\label{H0}
H_m = &J \sum_{\langle i,j \rangle} \v S_i\cdot \v S_j - K \sum_i (S_{i,z})^2 - b \sum_i S_{i,z},
\end{align}
where the first and second term are the antiferromagnetic exchange interaction $(J > 0)$
and the easy-axis anisotropy $(K > 0)$, respectively.
The parameter $b (= g \mu_B B)$ in the last term represents the external magnetic field along $z$-direction, where $g$ is the the g-factor and $\mu_B$ is the Bohr magneton.
The nearest-neighbor vectors are ${\v a}_1 = a(1,0)$, ${\v a}_2 = \frac{a}{2}(-1,\sqrt3)$ and ${\v a}_3 = -\frac{a}{2}(1,\sqrt3)$.
The next-nearest-neighbor vector are defined as ${\v b}_1 = {\v a}_1 - {\v a}_3$, ${\v b}_2 = {\v a}_2 - {\v a}_1$, and ${\v b}_3 = {\v a}_3 - {\v a}_1$ (see Fig.~\ref{fig:1}).

A sufficiently strong magnetic field $B$ destabilizes the collinear AFM order, leading to a spin-flop phase transition.
For $B < B_{\rm sf}$, where
\begin{align}
B_{\rm sf} = \frac{2S\sqrt{(3J-K)K}}{g\mu_B},
\end{align}
the equilibrium spin configuration forms a collinear AFM state along the $z$-axis.
In the spin-flop phase ($B > B_{\rm sf}$), the spin directions of the two sublattices are canted along the field direction (see Fig.~\ref{fig:1}).
Below, we compute the magnon spectra for these two distinct equilibrium spin configurations.

\subsubsection{Magnon Hamiltonian in a collinear AFM state}

In the collinear AFM state with the spin configuration ${\v S}_{i\in A} = S \hat {\v z}$ and ${\v S}_{i\in B} = -S \hat {\v z}$,
we perform the Holstein-Primakoff transformation and take the Fourier transformation, leading to the magnon Hamiltonian:
\begin{align}
H_m = \frac12 \sum_{\v k} \psi^\dag_{m,\v k} {\cal H}_m({\v k})  \psi_{m,\v k},
\end{align}
where $\psi_{m,\v k} = \left(a_{\v k}, b_{\v k}, a^\dag_{-\v k} , b^\dag_{-\v k}\right)^T$ is the basis function and the momentum space Hamiltonian is given by:
\begin{equation}
\begin{aligned}
{\cal H}^m_{\v k} = JS\left(
                    \begin{array}{cccc}
                      3 + \kappa_+ & 0 & 0 &  f_{\v k}  \\
                      0 & 3 + \kappa_- &  f^\ast_{\v k}  & 0 \\
                      0 &  f_{\v k}  & 3 + \kappa_+  & 0 \\
                       f^\ast_{\v k}  & 0 & 0 & 3 + \kappa_- \\
                    \end{array}
                  \right),
\end{aligned}
\end{equation}
where $\kappa_\pm = (2K \pm b/S)/J$ and $f_{\v k} = \sum_{j} e^{i \v k \cdot \v a_{j}}$.
To diagonalize the Bogoluibov Hamiltonian, we find the paraunitary matrix $U_{\v k}$ satisfying $E_{\v k} = U^{\dag}_{\v k} {\cal H}_{\v k} U_{\v k} = {\rm diag}( \epsilon_{\v k}, \epsilon_{-\v k})$.
Following the notation used in Ref.~\cite{Li2020}, we write the eigenvalue equations
\begin{equation}\label{PEE}
\begin{aligned}
{\sigma_3} {\cal H}_{\v k} |u^R_{n,\v k}\rangle = \bar\epsilon_{n,\v k} |u^R_{n,\v k}\rangle,\\
\langle u^L_{n,\v k}| {\sigma_3} {\cal H}_{\v k}  = \bar\epsilon_{n,\v k} \langle u^L_{n,\v k}|,
\end{aligned}
\end{equation}
where ${\sigma_3} = \rm diag(1,1,-1,-1)$ is the Pauli matrix acting on the particle-hole space.
Here $\langle u^L_{n,\v k}| = \langle u^R_{n,\v k}|{\sigma_3} $ and $|u^R_{n,\v k}\rangle = U_{n,\v k}$ are the left- and right-eigenvectors of the pseudo-Hermitian Hamiltonian ${\sigma_3}  {\cal H}_{\v k}$, respectively.
The orthonormal relation of the eigenvectors reads $\langle u^L_{n,\v k}|u^R_{m,\v k}\rangle = \langle u^R_{n,\v k}|{\sigma_3} |u^R_{m,\v k}\rangle= ({\sigma_3} )_{nm}$ and
the pseudo-eigenvalue satisfies $\bar\epsilon_{n,\v k} = ({\sigma_3} \epsilon_{\v k})_{nn}$.
By solving Eq.~\eqref{PEE}, we determine the energy eigenvalues of the magnon bands:
\begin{align}
%\epsilon_{\v k}^{\alpha/\beta} = \epsilon_{\v k}^0 \pm (D g_{\v k} + g\mu_B B),
\epsilon_{m,\v k}^{{\rm AFM},\pm} = JS\sqrt{(3 + \kappa)^2 - |f_{\v k}|^2} \pm b,
\end{align}
where $\kappa = 2K/J$.
The two magnonic states, denoted as $-$ and $+$, carry opposite spin angular momenta, with $S_z = -1$ and $S_z = +1$, respectively.

\subsubsection{Magnon Hamiltonian in a spin-flop state}

In the spin-flop phase, the equilibrium spin configuration is
\begin{equation}
\begin{aligned}
&{\v S}_A = S(-\cos\theta, 0, \sin\theta),\\
&{\v S}_B = S(\cos\theta, 0, \sin\theta),
\end{aligned}
\end{equation}
where $\theta = \sin^{-1} \left(\frac{b}{2(3J-K)S}\right)$ is the canted angle.
Here we assume the equilibrium spin profile lies in the $xz$-plane, for simplicity.
To investigate the spin-wave dynamics in the canted AFM phase, we introduce the sublattice dependent coordinate system (${\v S'}_{\alpha}$)
where the equilibrium spin direction is along the $z$-axis~\cite{Ganesh2011}
\begin{align}\label{SDtr}
{\v S}_{\alpha} = \left(
        \begin{array}{ccc}
          \sin\theta & 0 & (-1)^{\nu+1}\cos\theta \\
          0 & 1 & 0 \\
          (-1)^{\nu}\cos\theta & 0 & \sin\theta \\
        \end{array}
      \right) {\v S'}_{\alpha},
\end{align}
where $\nu = 1,2$ is the sublattice index ($A =1, B = 2)$.
By substituting the new spin variables into Eq.~\eqref{H0}, and neglecting the magnon-magnon interaction terms,
we obtain the low-energy Hamiltonian describing linear spin waves~\cite{Owerre2016}
\begin{equation}
\begin{aligned}
H = & -J \sum_{\langle i,j \rangle} \left[\cos2\theta ({S'}^x_i {S'}^x_j +  {S'}^z_i {S'}^z_j) - {S'}^y_i {S'}^y_j \right]\\
&- K \sum_i \left[ ({S'}^z_i)^2 \sin^2\theta + ({S'}^x_i)^2 \cos^2\theta\right] -  b \sum_i {S'}^z_i.
\end{aligned}
\end{equation}
The Holstein-Primakoff approach yields the Bogoliubov Hamiltonian
%\begin{align}
%H_m = \frac12 \sum_{\v k} \psi^\dag_\v k {\cal H}_{\v k} \psi_\v k, \qquad \psi_\v k = \left(\frac{a_{\v k} + a^\dag_{-\v k} }{\sqrt2},
%\frac{b_{\v k} + b^\dag_{-\v k} }{\sqrt2}, \frac{a_{\v k} - a^\dag_{-\v k} }{\sqrt2 i}, \frac{b_{\v k} - b^\dag_{-\v k} }{\sqrt2 i}\right)^T,
%\end{align}
%with
\begin{align}
{\cal H}^m_{\v k} = \left(
                    \begin{array}{cccc}
                      I_K            & F_{1,\v k}   & K_0            &  F_{2,\v k} \\
                      F^\ast_{1,\v k}   & I_K        & F^\ast_{2,\v k}  & K_0 \\
                      K_0             &  F_{2,\v k}  & I_K            &  F_{1,\v k} \\
                      F^\ast_{2,\v k}   & K_0        & F^\ast_{1,\v k}  & I_K\\
                    \end{array}
                  \right),
\end{align}
where
\begin{equation}
\begin{aligned}
&I_K = 3 J S \cos2\theta + K S (2\sin^2\theta - \cos^2\theta) + b \sin\theta,\\
&F_{1,\v k} =  J S f_{\v k} \sin^2\theta,\qquad F_{2,\v k} = - J S f_{\v k} \cos^2\theta,\\
&K_0 = -K S \cos^2\theta.
\end{aligned}
\end{equation}
By solving the eigenvalue equation in the particle-hole space, we obtain the magnon band dispersion
\begin{align}
&\epsilon_{m,\v k}^{{\rm SF},\pm} = \sqrt{(I_K^- \pm J S |f_{\v k}|)(I_K^+ \mp J S |f_{\v k}|\cos2\theta}),
\end{align}
where $I_K^\pm = I_K \pm K_0$.
It is noteworthy that, the azimuthal angle of the equilibrium state is arbitrarily chosen in the spin-flop phase,
leading to the breaking of U(1) spin-rotational symmetry.
For $\v k = 0$, the low-energy magnonic mode $\epsilon^{{\rm SF},-}_{m,\v k = 0} = 0$ is the gapless mode associated with spontaneous breaking of the U(1) symmetry,
while the high-energy mode $\epsilon^{{\rm SF},+}_{m,\v k = 0} \approx \sqrt{b^2 - 12 J K S^2}$ corresponds to the quasi-ferromagnetic resonance mode~\cite{Foner1963,Li2020b}.

\subsection{Phonon part}

The elastic Hamiltonian describing the lattice dynamics can be written as
\begin{align}\label{Ham0}
H_{p} = \sum_i \frac{\v p_i^2}{2M} + \frac{1}{2} \sum_{i,j,\alpha,\beta} u_i^\alpha \Phi_{i,j}^{\alpha,\beta} u_j^\beta,
\end{align}
where ${\v u}_i$ is the displacement vector of the $i$th ion from its equilibrium position,
$\v p_i$ is the conjugate momentum vector, $M$ is ion mass, and $\Phi_{i,j}^{\alpha,\beta}$ is a force constant matrix.
By using the Fourier transformation, we obtain the momentum space Hamiltonian
\begin{align}\label{Ham1}
H_{p} = \sum_{\v k} \left[\frac{{p}^\alpha_{-\v k} {p}^\alpha_{\v k}}{2M} + \frac{1}{2} {u}^\alpha_{-\v k} \Phi(\v k)^{\alpha \beta} {u}^\beta_{\v k}\right].
\end{align}
By introducing the dimensionless variables ${\bar u}^\alpha_{\v k} = \sqrt{\frac{M\omega_0}{\hbar}} {u}^\alpha_{\v k}$ and
${\bar p}^\alpha_{\v k} = \frac{1}{\sqrt{M\hbar\omega_0}} {p}^\alpha_{\v k}$, we rewrite
the Hamiltonian \eqref{Ham1} as
\begin{align}
H_{p} = \frac{1}{2} \sum_{\v k} \psi_{p,\v k}^\dag {{\cal H}^{p}_{\v k}} \psi_{p,\v k},
\end{align}
where the momentum space Hamiltonian is
\begin{align}
{{\cal H}^{p}_{\v k}} = \left(
                    \begin{array}{cc}
                      \frac{\hbar}{M\omega_0}\Phi(\v k) & 0 \\
                      0 & \hbar \omega_0 I_{6\times6} \\
                    \end{array}
                  \right).
\end{align}
and the basis function is
\begin{widetext}
\begin{align}
\psi_{p,\v k}= ({\bar u}^A_{x,\v k}, {\bar u}^A_{y,\v k}, {\bar u}^A_{z,\v k}, {\bar u}^B_{x,\v k}, {\bar u}^B_{y,\v k}, {\bar u}^B_{z,\v k}, {\bar p}^A_{x,\v k}, {\bar p}^A_{y,\v k}, {\bar p}^A_{z,\v k}, {\bar p}^B_{x,\v k}, {\bar p}^B_{y,\v k}, {\bar p}^B_{z,\v k})^T, \qquad \psi_{p,\v k}^\dag = \psi_{p,-\v k}.
\end{align}
\end{widetext}
The momentum space representation of the force constant matrix is expressed as
\begin{align}
\Phi({\v k}) = \left(
        \begin{array}{cc}
          \Phi^{AA} ({\v k}) & \Phi^{AB} ({\v k}) \\
          \Phi^{BA} ({\v k}) & \Phi^{BB} ({\v k}) \\
        \end{array}
      \right),
\end{align}
where $\Phi^{AB}({\v k}) = \Phi^{BA}(-{\v k})$ represents the elastic interaction between different sublattices from the nearest-neighbor couplings.
In the first- and second-neighbor approximation of the 2D honeycomb lattice,
the in-plane and out-of-plane phonon modes do not couple~\cite{Falkovsky2007}, and the momentum space force constant matrix between the nearest-neighbors is given by
\begin{align}\label{eqPhiAB}
\Phi^{AB}_{ij}(\v k) =&\, \Phi^{AB}_{ij}({\v a}_1) e^{i k_x a} + \Phi^{AB}_{ij}({\v a}_2) e^{i (-\frac{k_x a}{2} + \frac{\sqrt{3} k_y a}{2} ) }\nn
&+  \Phi^{AB}_{ij}({\v a}_3) e^{i (-\frac{k_x a}{2} - \frac{\sqrt{3} k_y a}{2} ) },
\end{align}
where $i$ and $j$ represent phonon modes in cartesian coordinates, i.e., $\{i,j\} \in (x,y,z)$.
The first component of the force constant matrix in Eq.~\eqref{eqPhiAB} is given by
\begin{align}\label{FCM1}
\Phi^{AB}_{ij}({\v a}_1) = \left(
                                          \begin{array}{ccc}
                                            K_L & 0 & 0\\
                                            0 & K_T & 0\\
                                            0 & 0 & K_Z\\
                                          \end{array}
                                        \right).
\end{align}
The other components are obtained by the $C_3$ rotation around $z$-axis~\cite{Falkovsky2007}
\begin{align}
\Phi^{AB}_{ij}({\v a}_m) = U_m^{-1} \Phi^{AB}_{ij}({\v a}_1) U_m, \qquad (m=2,3)
\end{align}
where
\begin{align}
U_m = \left(
        \begin{array}{ccc}
          \cos\theta_m & \sin\theta_m & 0\\
          -\sin\theta_m & \cos\theta_m & 0\\
           0 & 0 & 1\\
        \end{array}
      \right),
\end{align}
with $\theta_2 = \frac{2\pi}{3}$ and $\theta_3 = -\frac{2\pi}{3}$.
Therefore, we have
\begin{align}
\Phi^{AB}({\v k}) = \left(
        \begin{array}{ccc}
          \Phi^{AB}_{xx} ({\v k}) & \Phi^{AB}_{xy} ({\v k}) & 0 \\
          \Phi^{AB}_{yx} ({\v k}) & \Phi^{AB}_{yy} ({\v k}) & 0 \\
          0 & 0 & \Phi^{AB}_{zz} ({\v k}) \\
        \end{array}
      \right),
\end{align}
where
\begin{align}
&\Phi^{AB}_{xx} ({\v k}) = K_L e^{i k_x} + \frac{1}{2}(K_L + 3K_T) e^{-\frac{i k_x}{2}}\cos\left(\frac{\sqrt3}{2} k_y\right),\nn
&\Phi^{AB}_{xy} ({\v k}) = \Phi^{AB}_{yx} ({\v k}) = i \frac{\sqrt3}{2} (K_T - K_L) e^{-\frac{i k_x}{2}}\sin\left(\frac{\sqrt3}{2} k_y\right),\nn
&\Phi^{AB}_{yy} ({\v k}) = K_T e^{i k_x} + \frac{1}{2}(3K_L + K_T) e^{- \frac{i k_x}{2}}\cos\left(\frac{\sqrt3}{2} k_y\right),\nn
&\Phi^{AB}_{zz}(\v k) = K_Z f_{\v k}.
\end{align}
The diagonal elements of the force constant matrices are
\begin{align}
\Phi^{AA} ({\v k}) = \Phi^{BB} ({\v k}) = 3\left(
                                          \begin{array}{ccc}
                                            \frac{K_L + K_T}{2} & 0 & 0\\
                                            0 & \frac{K_L + K_T}{2} & 0\\
                                            0 & 0 & K_Z\\
                                          \end{array}
                                        \right),
\end{align}
which ensures that the lowest energy of the acoustic phonon modes remains zero.

\subsection{Magnon-phonon interaction Hamiltonian}

The general expression of the magnetoelastic interaction is given by~\cite{Kittel1949, Kittel1958}
\begin{align}\label{Hmp}
H_{mp} &= - \sum_i {\v S}^T_i E_i {\v S}_i = - \sum_\nu \sum_{i\in \nu} {\v S}^T_{\nu,i} E_{i}^\nu {\v S}_{\nu,i},
\end{align}
where $\nu = 1,2$ is the sublattice index ($A =1, B = 2)$.
The coupling matrix is written in terms of the cartesian strain tensor:
\begin{align}
&\epsilon^{i}_{\alpha \beta} = \frac{1}{2}\left(\frac{\partial u^\beta_i}{\partial r_\alpha} + \frac{\partial u^\alpha_i}{\partial r_\beta}\right).
\end{align}
In the hexagonal systems, the magnetoelastic Hamiltonian in Ref.~\cite{Callen1965} leads to
\begin{align}
&E_i^\nu = \frac12\left(
        \begin{array}{ccc}
          B^\gamma \epsilon^{\gamma}_1   & B^\gamma\epsilon^{\gamma}_2 & B^\epsilon\epsilon^{\epsilon}_2 \\
          B^\gamma\epsilon^{\gamma}_2 & -B^\gamma \epsilon^{\gamma}_1 & B^\epsilon\epsilon^{\epsilon}_1 \\
          B^\epsilon\epsilon^{\epsilon}_2 & B^\epsilon\epsilon^{\epsilon}_1 &\sqrt3\left(B^\alpha_{12} \epsilon^{\alpha}_1 + B^\alpha_{22}\epsilon^{\alpha}_2 \right)\\
        \end{array}
      \right),
\end{align}
where $\epsilon^{\alpha}_1 = \epsilon_{xx} + \epsilon_{yy}+ \epsilon_{zz}$, $\epsilon^{\alpha}_2 = \frac{\sqrt3}{2}\left(\epsilon_{zz} - \frac{1}{3}\epsilon^{\alpha}_1\right)$,
$\epsilon^{\gamma}_1 = \frac{1}{2}\left(\epsilon_{xx} - \epsilon_{yy} \right)$, $\epsilon^{\gamma}_2 = \epsilon_{xy}$, $\epsilon^{\epsilon}_1 = \epsilon_{yz}$,
and $\epsilon^{\epsilon}_2 = \epsilon_{xz}$ are the symmetrized strains in the hexagonal point group.
In a discrete system, the local strains can be written as~\cite{Holm2018, Huang2021}:
\begin{align}
\epsilon_{\alpha\beta}^i \rightarrow \bar\epsilon_{\alpha\beta}^{i,\nu} = \frac{1}{n} \sum_{m=1}^3 \bar\epsilon_{\alpha\beta}^{i,i+{{\v a}^D_m}},
\end{align}
where ${\v a}_m^\nu$ is the nearest-neighbor vector (${\v a}_m^A = -{\v a}_m^B$), and the
components of the strain tensor are:
\begin{align}
\bar\epsilon_{\alpha\beta}^{i,i+{{\v a}_m^\nu}}% &= \epsilon_{\beta \alpha}^{i,i+{{\v a}_m^D}}\nn
= \frac{1}{2} \left[a_m^{\nu,\alpha} (u_i^\beta - u_{i+{{\v a}_m^\nu}}^\beta) + a_m^{\nu,\beta} (u_i^\alpha - u_{i+{{\v a}_m^\nu}}^\alpha)\right],
\end{align}
where $n$ is a normalization constant that ensures $\bar\epsilon_{\alpha\beta}^{i,\nu}$ reduced to $\epsilon_{\alpha\beta}^i$ in the long-wavelength limit (in honeycomb lattice $n = -3a^2/2$).
The explcit expression for the local strain components is given by
\begin{equation}\label{eqLC}
\begin{aligned}
&\bar \epsilon_{xx}^{i,A} = -\frac{1}{3a}  \left( - 2 u^x_{B,i + {\v a}_1} + u^x_{B,i + {\v a}_2} + u^x_{B,i + {\v a}_3}\right),\\
&\bar \epsilon_{yy}^{i,A} = -\frac{1}{\sqrt3 a}  \left( -u^y_{B,i + {\v a}_2} + u^y_{B,i + {\v a}_3}\right),\\
&\bar \epsilon_{xy}^{i,A} = -\frac{1}{6a}  \left( - 2u^y_{B,i + {\v a}_1} + u^y_{B,i + {\v a}_2} + u^y_{B,i + {\v a}_3}\right.\\
&\hspace{18mm} \left.- \sqrt3 u^x_{B,i + {\v a}_2} + \sqrt3 u^x_{B,i + {\v a}_3}\right),\\
&\bar \epsilon_{xz}^{i,A} = -\frac{1}{6a}  \left( - 2u^z_{B,i + {\v a}_1} + u^z_{B,i + {\v a}_2} + u^z_{B,i + {\v a}_3} \right),\\
&\bar \epsilon_{yz}^{i,A} = -\frac{1}{2\sqrt3 a} \left( - u^z_{B,i + {\v a}_2} + u^z_{B,i + {\v a}_3} \right).
\end{aligned}
\end{equation}
Here, $\epsilon_{\alpha\beta}^{i,B}$ can be obtained by replacement $B \rightarrow A$ and $\v a_i \rightarrow - \v a_i$.
The explicit expressions of the quadratic Hamiltonian describing the magnon-phonon interaction that includes a magnon operator and a phonon operator are shown in Appendix~\ref{secA1}.

\begin{figure*}[t!]
\includegraphics[width=2.0\columnwidth]{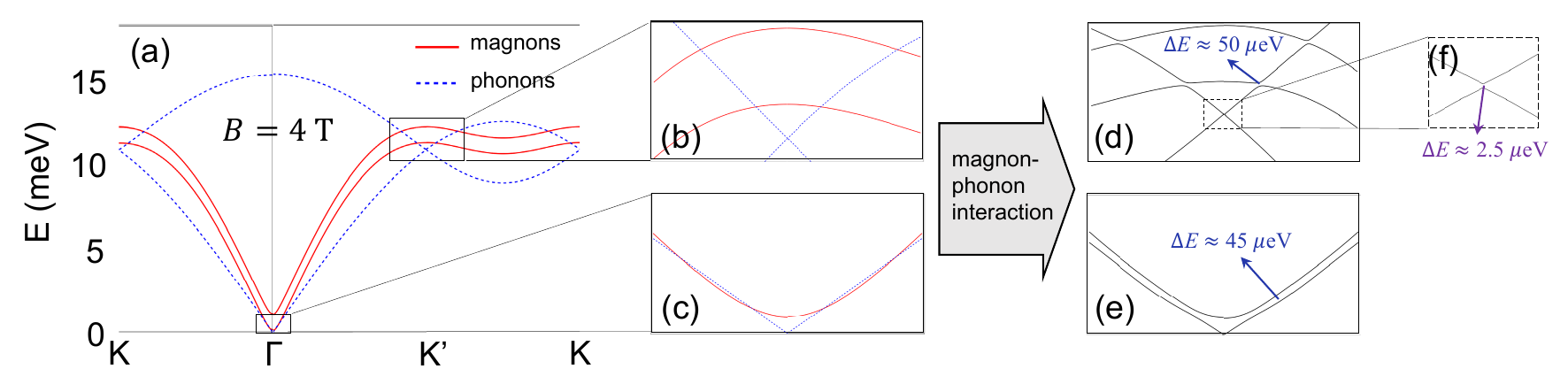}
\caption{Band structures of magnon and phonon modes in a collinear antiferromagnetic (AFM) state (a-c) in the absence of magnon-phonon interaction and (d-f) in the presence of magnon-phonon interaction.
A barely visible band gap between two phonon-like modes is numerically confirmed in (f).
Here, we use $B^\epsilon =-0.84$~meV and $B = 4$~T.}\label{fig:2}
\end{figure*}
\begin{figure*}[t!]
\includegraphics[width=2.0\columnwidth]{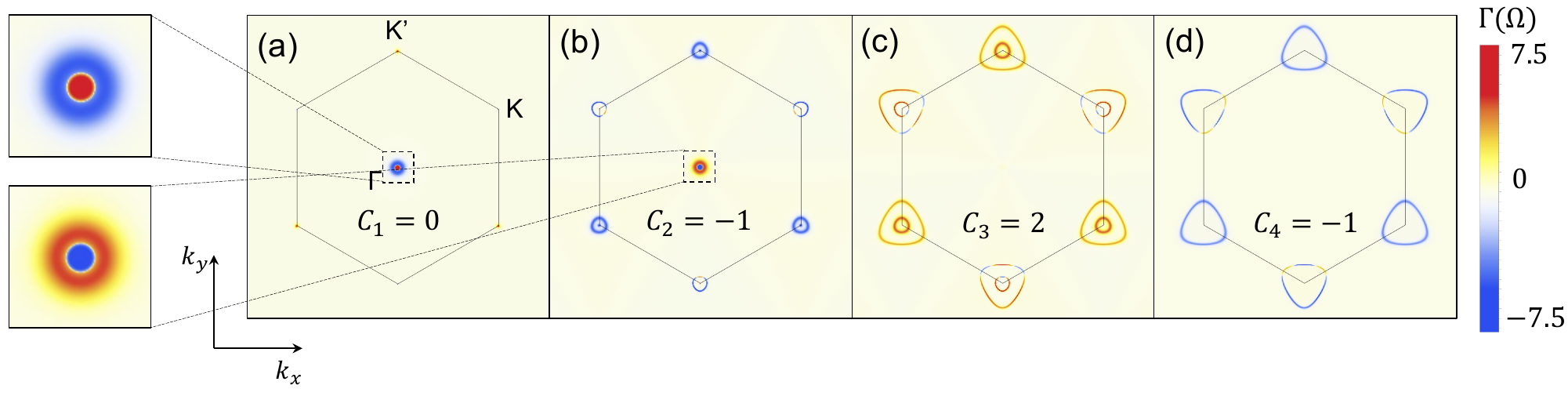}
\caption{Profiles of the Berry curvature in log scale $\Gamma(\Omega) = {\rm sgn}(\Omega) {\rm log} (1 + |\Omega|)$. The hexagon represents the 1st Brillouin zone. Here, we use $B^\epsilon =-0.84$~meV and $B = 4$~T.}\label{fig:3}
\end{figure*}

\section{Band topology of magnon-polarons}

Here, we present the computational results for the band structure of the magnon-polaron bands and the corresponding topological transport properties of the magnon-polarons.
{In this section, we utilize the magnetic parameter of MnPS$_3$ as it is known to exhibit a field-induced spin-flop transition at around 5 T~\cite{Goossens2000}: $J = 1.54$ meV, $S = 5/2$~\cite{Wildes1998} and $K S = 7.25$ $\mu$eV.
For the phonons, we use following parameters. The lattice constant is $a \approx 3.5$ {\AA}~\cite{Ouvrard1985}, Mn ion mass is $M = 55$ u, and the spring constants are taken in agreement with the phonon spectrum of MnPS$_3$: $K_L = 300$ eV/nm$^2$, $K_L = 110$ eV/nm$^2$ and $K_Z = 50$ eV/nm$^2$~\cite{Yang2020}.
For the magnetoelastic constants, we use $B^\epsilon = B^\alpha_{12} = B^\alpha_{22} = B^\gamma = - 0.84$ meV, resulting in $B^\epsilon/V \approx-1.2 \times 10^7$ erg/cm$^3$, which is comparable to Kittel's estimation for iron~\cite{Kittel1949}.

The magnon-polaron Hamiltonian in our model is expressed as
\begin{align}
H = H_{m} + H_{p} + H_{ph} = \frac{1}{2} \sum_{\v k} \psi^\dag_{\v k} {\cal H}_{\v k} \psi_{\v k}.
\end{align}
Due to the complexity of our model Hamiltonian, which is a $16\times 16$ matrix, analytical computation of the eigenvalue problem is challenging.
In this context, we present numerical results for the band structure and the topological properties of the magnon-polarons.

The band topology is characterized by the Berry curvature of the Bogoluibov Hamiltonian~\cite{Li2020},
\begin{align}
&\Omega^n_z ({\v k}) = (\sigma_3)_{nn}(\sigma_3)_{mm} \sum_{n\neq m} \frac{-2 {\rm Im}[\langle n|v_x |m\rangle\langle m|v_y |n\rangle]}{(\epsilon^n_{\v k} - \epsilon^m_{\v k})^2 + \delta^2},
\end{align}
where $v_i = \frac{1}{\hbar}\frac{\partial {\cal H}_{\v k}}{\partial \v k}$ is the velocity operator, $|n\rangle$ and $\epsilon^n_{\v k}$ are the $n$-th right-eigenvector and corresponding eigenvalue of the Bogoluibov Hamiltonian, respectively.
Here, we introduce a level broadening parameter $\delta$, set to be $0$ in this section. We consider finite level broadening in the thermal Hall transport calculation in Section~\ref{sec4}.
In this section, we employ the numerical algorithm proposed by Fukui, Hatsugai and Suzuki~\cite{Fukui2005} to efficiently compute the Chern number of the magnon-polaron bands.

\subsubsection{Band topology in a collinear AFM state}

In the collinear AFM state, the magnon-phonon interaction excludes the in-plane phonon amplitudes~\cite{Zhang2020}.
As the band topology remains unaffected by these in-plane phonon modes, we simplify the model by disregarding these modes.
Consequently, we consider an $8\times8$ Bogoliubov Hamiltonian that includes 2 magnon modes ($a_{\v k}$ and $b_{\v k}$)
and 2 out-of-plane phonon modes ($\bar u^A_{z,\v k}$ and $\bar u^B_{z,\v k}$) for the particle and hole sectors, respectively.

In Figure~\ref{fig:2}, we depict the band structures for the collinear AFM phase in the presence of a perpendicular magnetic field ($B = 4$ T),
both with and without magnon-phonon interaction.
With finite magnon-phonon interaction, anticrossing band gaps emerge between magnon and phonon modes near $\Gamma$- and $K$$(K')$-points [Figs.~\ref{fig:2}(d) and (e)].
These gaps lead to the formation of nodal rings in the Brillouin zone, indicating the presence of topological magnon-polarons around these points.

Because there are two anticrossings near $\v k =0$ ($\Gamma$) which exhibit opposite topologies [see Figs.~\ref{fig:2}(c) and (e)],
a sign change of the Berry curvature is observed near $\Gamma$-point in the lowest two bands [see Figs.~\ref{fig:3}(a) and (b)].
At low temperatures, the band topology near $\v k =0$ contributes significantly to the thermal Hall transport, as detailed in Section~\ref{sec4}.
Additionally, the Berry curvatures remain finite in the vicinity of the $K(K')$-point due to the nodal ring at these locations.
It's noteworthy that the magnon-phonon interaction introduces a slight band gap between two phonon-like bands at the $K(K')$-point
[Figs.~\ref{fig:2}(f)], inducing a large Berry curvature at this point.
However, the band topology at this slight anticrossing gap does not affect the Hall transport due to the suppressing effect of the finite level broadening.

\begin{figure*}[t!]
\includegraphics[width=2.0\columnwidth]{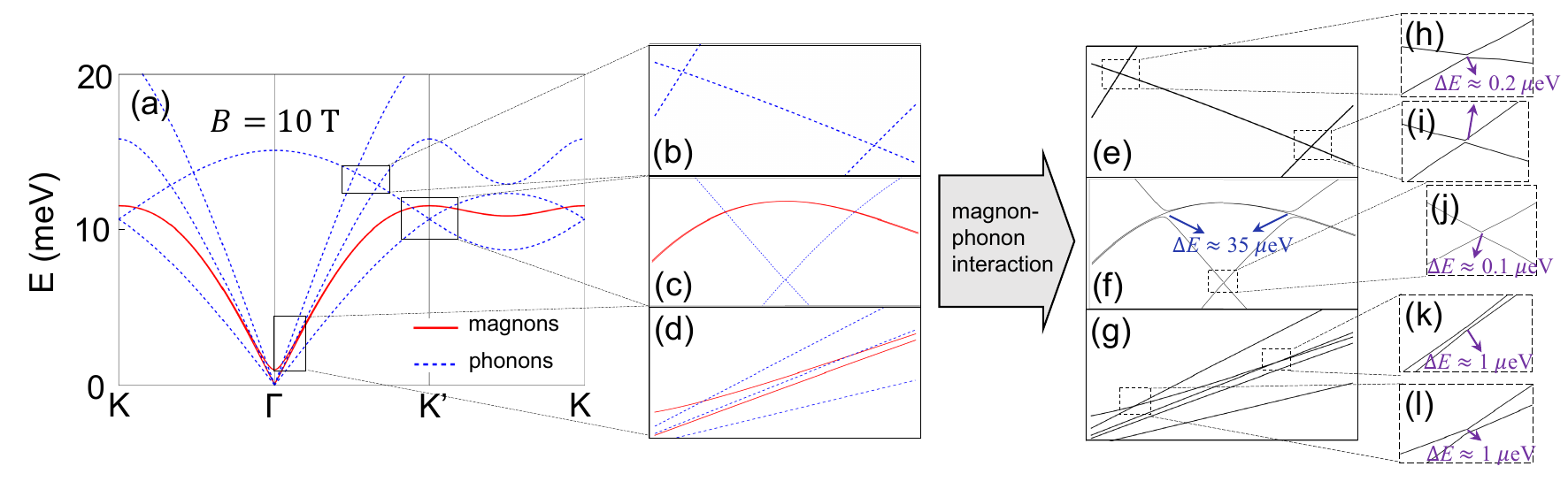}
\caption{Band structures of magnon and phonon modes in the spin-flop state (a-d) in the absence of magnon-phonon interaction and (e-l) in the presence of magnon-phonon interaction.
Barely visible band gaps are numerically confirmed in (h-l).
Here, we use $B^\epsilon = B^\alpha_{12} = B^\alpha_{22} = B^\gamma = - 0.84$ meV and $B = 10$~T.}\label{fig:4}
\end{figure*}

\begin{figure*}[t!]
\includegraphics[width=2.1\columnwidth]{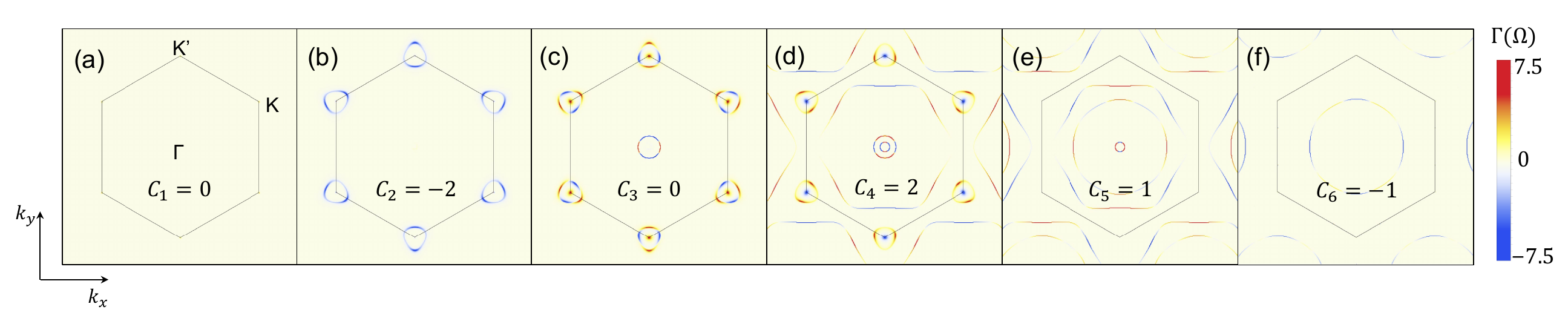}
\caption{Profiles of the Berry curvature in log scale $\Gamma(\Omega) = {\rm sgn}(\Omega) {\rm log} (1 + |\Omega|)$ in the spin-flop state.
The hexagon represents the 1st Brillouin zone.
Here, we use $B^\epsilon = B^\alpha_{12} = B^\alpha_{22} = B^\gamma = - 0.84$ meV and $B = 10$~T.}\label{fig:5}
\end{figure*}

\subsubsection{Band topology in a spin-flop state}

In the spin-flop state, the magnon-phonon interaction involves both in-plane and out-of-plane phonon amplitudes.
In this case, we consider a $16\times 16$ Bogoliubov Hamiltonian including 2 magnon modes and 6 phonon modes for the particle and hole sectors, respectively.
In Fig.~\ref{fig:4}, we show the low-energy band structures for the spin-flop phase under the influence of a perpendicular magnetic field $B = 10$~T, both with and without magnon-phonon interaction.

Because $b \ll 6 J S$ for $B = 10$~T, the spin canted angle is small ($\theta \approx 0.05$), and there is also a small band gap between the two magnonic bands.
Similar to the collinear AFM case, the magnon-phonon interaction induces anticrossing gaps between magnon and phonon modes.
In the spin-flop state, the gapless magnonic mode and the acoustic phonon modes converge at $\v k = 0$, and the magnon-polaron band topology is not well-defined at this point.
To establish a well-defined band topology theoretically, we introduce a sufficiently small easy-axis magnetic anisotropy along the bond direction, which should be present due to the lattice structure, but is higher-order anisotropy than the other terms in our magnetic Hamiltonian and therefore expected to induce negligible physical effects.
Close to the $\Gamma$- and $K$-points, the magnonic mode couples with the phonon modes, resulting in the formation of anticrossing gaps~[Figs.~\ref{fig:4}(c,d) and (f,g,k,l)].
Additionally, we numerically confirm the slight energy gaps between two phonon-like modes [Figs.~\ref{fig:4}(h-j)].
Here, we omit the in-plane optical phonon modes, because these high-energy phonons do not couple with the magnons, displaying trivial topology ($C_7 = C_8 = 0$).

In Fig.~\ref{fig:5}, we present the Berry curvatures and corresponding Chern numbers of the lowest 6 modes.
Figure~\ref{fig:4} and \ref{fig:5} illustrate one of our main findings: the exploration of the topological characteristics of the magnon-polaron,
incorporating both in-plane and out-of-plane phonon modes in the AFM spin-flop state, which has not been addressed in previous studies.
It is noteworthy that, in contrast to the collinear AFM case, the magnon-polaron gaps near the $\v k = 0$ are very small~[Figs.~\ref{fig:4}(g) and (k,l)].
Therefore, we anticipate a small thermal Hall effect at low temperatures in the spin-flop state.

\section{Thermal Hall effect}\label{sec4}

The topology of magnon-phonon hybrid excitations gives rise to the intrinsic thermal Hall effect.
The Berry-curvature-induced thermal Hall conductivity is given by~\cite{Matsumoto2011,Matsumoto2014a}
\begin{align}
\kappa_{xy}^{2D} = -\frac{k_B^2 T}{\hbar V} \sum_{n, \v k} \left[c_2(\rho_{n,\v k}) - \pi^2/3\right] \Omega_z^n(\v k),
\end{align}
where $c_2(\rho) = (1+\rho) \ln^2 [(1+\rho) / \rho] - \ln^2 \rho - 2 {\rm Li}_2(-\rho)$, $\rho_{n,\v k} = (e^{E_n(\v k)/{k_B T}} - 1)^{-1}$ is the Bose-Einstein distribution function with a zero chemical potential,
$k_B$ is the Boltzmann constant, $T$ is the temperature, and ${\rm Li}_2 (z)$ is the polylogarithm function.
Here, we use the same parameters used in the previous section except for a constant level broadening parameter $\delta = 1$~$\mu$eV.
To convert two-dimensional conductivity to bulk conductivity, we employ the relation $\kappa_{xy} = \kappa_{xy}^{2D}/d$, with $d = 0.67$~nm representing the monolayer thickness of MnPS$_3$~\cite{Ouvrard1985,Lim2021}.

In Fig.~\ref{fig:6}, a key outcome of our study, we illustrate the magnetic field and temperature dependence of the thermal Hall conductivity for magnon-polarons, considering both in-plane and out-of-plane lattice dynamics.
The thermal Hall conductivity exhibits an increase with the magnetic field, changing sign at the phase transition point~[Fig.~\ref{fig:6}(a)], in agreement with the results reported in Ref.~\cite{Li2023}.
In the collinear AFM phase, the thermal Hall conductivity is negative at low temperatures but undergoes a sign change as temperature increases.
This behavior can be attributed to the sign change in the Berry curvature near the $\Gamma$ point of the lowest modes [Fig.~\ref{fig:3}(a) and (b)].
It is worth noting that, owing to the small anticrossing gap of the spin flop phase in the low-energy regime~[Figs.~\ref{fig:4}(g) and (k,l)],
the magnitude of the thermal Hall conductivity in the spin flop phase is significantly smaller than that in the collinear AFM phase,
as observed in our parameter setup based on MnPS$_3$.

\begin{figure*}[t!]
\includegraphics[width=1.6\columnwidth]{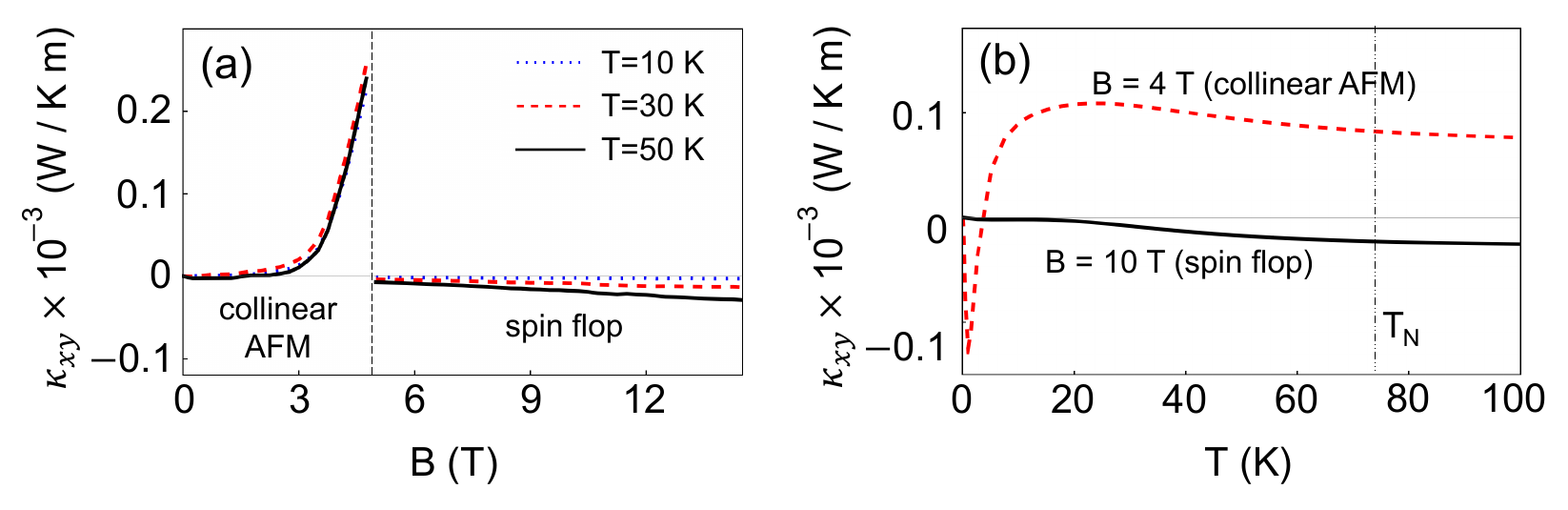}
\caption{Thermal Hall conductivity as a function of (a) an external magnetic field and (b) temperature. $T_N$ denotes the N\'{e}el temperature of the MnPS$_3$.
Although strictly speaking, magnons are defined well only below the Neel temperature, The calculation result is shown also above the Neel temperature to provide the qualitative temperature dependence of the thermal Hall conductivity for a broad range of temperatures.}\label{fig:6}
\end{figure*}

\section{Discussion}

In this study, we explore the topological properties of magnon-polaron bands in a two-dimensional honeycomb antiferromagnetic insulator by varying an external magnetic field.
We specifically investigate the impact of strain-dependent magnetic anisotropy, a common feature in magnetic materials, on the topological magnon-polarons.
The magnetic phase transition from the collinear antiferromagnetic phase to the spin-flop phase induces a notable transformation in the magnon-polaron band structure and its associated topological properties.
Consequently, the thermal Hall conductivity undergoes a change in sign and amplitude at the phase transition point.
Our findings provide insights for interpreting experimental data on thermal Hall conductivity in two-dimensional antiferromagnetic materials.

\begin{acknowledgments}
This work was supported by the Brain Pool Plus Program through the National Research Foundation of Korea funded by the Ministry of Science and ICT (2020H1D3A2A03099291) and the National Research Foundation of Korea funded by the Korea Government via the SRC Center for Quantum Coherence in Condensed Matter (RS-2023-00207732).
G.G. acknowledges support by the National Research Foundation of Korea (NRF-2022R1C1C2006578).
The work at SNU was supported by the Leading Researcher Program of Korea’s National Research Foundation (Grant No. 2020R1A3B2079375).
\end{acknowledgments}

\bibliography{reference_SF}

\appendix
\counterwithin{figure}{section}
\begin{widetext}
\section{Magnon-phonon interaction}\label{secA1}

\subsection{Magnon-phonon interaction in collinear AFM state}

In the collinear AFM state, off-diagonal components of the strain tensor involving out-of-plane phonon vibration only contributes to the magnetoelastic interaction to linear order in magnon amplitude.
Explicitly, we have%~\cite{Thingstad2019}
\begin{align}\label{Hmp2}
H_{mp} = -  \frac{B^\epsilon}{2}\sum_\nu \sum_{i\in \nu} S_\nu^{z} \Big(&S^x_\nu \epsilon^\nu_{xz}  + S^y_\nu \epsilon^\nu_{yz}
+\epsilon^\nu_{zx} S^x_\nu + \epsilon^\nu_{zy} S^y_\nu \Big)
\end{align}
By using Eq.~\eqref{eqLC} and taking the Fourier transformation with the Holstein Primakoff approach, we obtain
\begin{align}
H_{mp} &= \frac{\kappa_\epsilon}{2} \sum_{\v k} \Bigg[s_x^\ast {\bar u}^B_{z,-\v k}\left( \frac{a_{\v k} +  a^\dag_{-\v k}}{\sqrt2}\right) +  s_y^\ast {\bar u}^B_{z,-\v k}\left( \frac{a_{\v k} -  a^\dag_{-\v k}}{\sqrt2 i}\right)+ s_x {\bar u}^A_{z,-\v k}\left( \frac{b_{\v k} +  b^\dag_{-\v k}}{\sqrt2}\right) - s_y {\bar u}^A_{z,-\v k}\left( \frac{b_{\v k} -  b^\dag_{-\v k}}{\sqrt2 i}\right)\Bigg] + {\rm h.c.},
\end{align}
where $s_x = (-2e^{i \v k \cdot {\v a}_1 } + e^{i \v k \cdot {\v a}_2 } + e^{i \v k \cdot {\v a}_3 })$,
$s_y = \sqrt3  (-e^{i \v k \cdot {\v a}_2 } + e^{i \v k \cdot {\v a}_3 })$ (in the low $\v k$ limit, $s_x \approx -3 i k_x, s_y \approx -3 i k_y$)
and $\kappa_\epsilon = \frac{B^\epsilon S}{6 a}\sqrt{\frac{\hbar S}{M\omega_0}}$.
We note that the magnon-phonon interaction does not involve the in-plane phonon modes in the collinear AFM state.

\subsection{Magnon-phonon interaction in spin-flop state}

In the spin-flop phase, the sub-lattice dependent spin transformation~\eqref{SDtr} leads to
\begin{align}\label{Hmp2}
H_{mp} &= - S \sum_\nu \sum_{i\in \nu} (-1)^{\nu+1} \cos\theta \sin\theta \left[(E^\nu_{xx} - E^\nu_{zz}) {S'}^x_\nu + {S'}^x_\nu (E^\nu_{xx} - E^\nu_{zz}) \right] + (-1)^{\nu+1} \cos\theta (E^\nu_{xy} {S'}^y_\nu + {S'}^y_\nu E^\nu_{xy})\nn
&\hspace{21mm}  + (\sin^2\theta-\cos^2\theta ) (E^\nu_{xz} {S'}^x_\nu + {S'}^x_\nu E^\nu_{xz}) + \sin\theta ({S'}^y_\nu E^\nu_{yz} + E^\nu_{yz} {S'}^y_\nu)
\end{align}
In this case, the magnon-phonon interaction includes full in-plane and out-of-plane phonon modes.
In terms of dimensionless operators, the momentum space representation of the magnon-phonon interaction is
\begin{align}
H_{mp} =& \sin2\theta \sum_{\v k} \Bigg[\kappa_-  s_x^\ast u^x_{B,-\v k} \left(\frac{a_{\v k} + a^\dag_{-\v k}}{\sqrt2}\right)
+ \kappa_+ s_y^\ast u^y_{B,-\v k} \left(\frac{a_{\v k} + a^\dag_{-\v k}}{\sqrt2}\right)\nn
&\hspace{18mm}- \kappa_-  s_x u^x_{A,-\v k} \left(\frac{b_{\v k} + b^\dag_{-\v k}}{\sqrt2}\right)
- \kappa_+  s_y
u^y_{A,-\v k} \left(\frac{b_{\v k} + b^\dag_{-\v k}}{\sqrt2}\right)\Bigg]\nn
& + \frac{\kappa_\gamma}{2} \cos\theta\sum_{\v k}
\Bigg\{\left(s_x^\ast u^y_{B,-\v k} + s_y^\ast u^x_{B,-\v k} \right) \left(\frac{a_{\v k} - a^\dag_{-\v k}}{\sqrt2 i}\right)
- \left(s_x u^y_{A,-\v k} + s_y u^x_{A,-\v k} \right)
\left(\frac{b_{\v k} - b^\dag_{-\v k}}{\sqrt2 i}\right)\Bigg\} \nn
&-\frac{\kappa_\epsilon}{2} \cos2\theta\sum_{\v k}
\left[s_x^\ast u^z_{B,-\v k}
\left(\frac{a_{\v k} + a^\dag_{-\v k}}{\sqrt2}\right)
+ s_x u^z_{A,-\v k} \left(\frac{b_{\v k} + b^\dag_{-\v k}}{\sqrt2}\right)\right]\nn
&+ \frac{\kappa_\epsilon}{2} \sin\theta\sum_{\v k}
\left[s_y^\ast  u^z_{B,-\v k}
\left(\frac{a_{\v k} - a^\dag_{-\v k}}{\sqrt2 i}\right) +s_y u^z_{A,-\v k} \left(\frac{b_{\v k} - b^\dag_{-\v k}}{\sqrt2 i}\right)\right]+ {\rm h.c.}.
\end{align}
where $\kappa_\gamma = \frac{B^\gamma S}{6 a}\sqrt{\frac{\hbar S}{M\omega_0}}$ and $\kappa_\pm = \frac{B_\pm S}{6 a}\sqrt{\frac{\hbar S}{M\omega_0}}$ with
$B_\pm = \frac{1}{4} (-2 \sqrt3 B^\alpha_{12} + B^\alpha_{22}
\pm B^\gamma)$.
\end{widetext}

\end{document}